\documentclass[showpacs,aps,preprint,prl]{revtex4}

\usepackage{epsfig}
\usepackage{graphicx}
\usepackage{dcolumn}
\usepackage{bm}

\let\jnfont=\rm
\def\NPB#1,{{\jnfont Nucl.\ Phys.\ B }{\bf #1},}
\def\PLB#1,{{\jnfont Phys.\ Lett.\ B }{\bf #1},}
\def\EPJC#1,{{\jnfont Eur.\ Phys.\ Jour.\ C }{\bf #1},}
\def\PRD#1,{{\jnfont Phys.\ Rev.\ D }{\bf #1},}
\def\PRL#1,{{\jnfont Phys.\ Rev.\ Lett.\ }{\bf #1},}
\def\MPLA#1,{{\jnfont Mod.\ Phys.\ Lett.\ A }{\bf #1},}
\def\JPG#1,{{\jnfont J.\ Phys.\ G}{\bf #1},}
\def\CTP#1,{{\jnfont Commun.\ Theor.\ Phys.\ }{\bf #1},}
\def\JHEP#1,{{\jnfont JHEP \ }{\bf #1},}
\def\NPPS#1,{{\jnfont Nucl.\ Phys.\ Proc.\ Suppl.\ }{\bf #1},}

\def\lsim{\mathrel{\mathpalette\oversim<}}
\def\gsim{\mathrel{\mathpalette\oversim>}}
\def\oversim#1#2{\lower0.5ex\vbox{\baselineskip0pt\lineskip0pt
  \lineskiplimit0pt\everycr{}\tabskip0pt
  \halign{$\mathsurround0pt #1\hfil##\hfil$\crcr #2\crcr\sim\crcr}}}

\begin{document}

\preprint{\parbox{1.2in}{\noindent arXiv:0901.1437}}

\title{ Experimental constraints on nMSSM and implications \\ on its phenomenology }

\author{Junjie Cao$^{1,2}$,  Heather E. Logan$^2$, Jin Min Yang$^3$  \vspace*{0.2cm}}

\affiliation{$^1$ Department of Physics, Henan Normal University,
                  Xinxiang 453007, China \\
$^2$ Ottawa Carleton Institute for Physics, Carleton University, Ottawa, K1S 5B6 Canada  \\
$^3$ Key Laboratory of Frontiers in Theoretical Physics, Institute of Theoretical Physics,
    Academia Sinica, Beijing 100190, China
    \vspace*{0.5cm}}

\begin{abstract}
We examine various direct and indirect experimental constraints on
the nearly minimal supersymmetric standard model (nMSSM) and obtain
the following observations:
(i) Current experiments stringently constrain
    the parameter space, setting a range of
     $1\sim 37$ GeV for the lightest neutralino $\tilde{\chi}^0_1$,
     $30 \sim 140$ GeV ($1\sim 250$ GeV)
     for the lightest CP-even (CP-odd) Higgs boson,
     and $1.5 \sim 10$ for $\tan\beta$;
(ii) To account for the dark matter relic density,
     besides the s-channel exchange of a Z-boson, the s-channel exchange of a
     light $A_1$ (the lightest CP-odd  Higgs boson)
     can also play an important role in LSP annihilation.
     Compared with the $Z$-exchange annihilation channel,
      the $A_1$ exchange channel is more favored by muon $g-2$ data and
       allows much broader regions for the parameters;
(iii) In a large part of the allowed parameter space the SM-like Higgs boson
       may dominantly decay to $\tilde{\chi}^0_1 \tilde{\chi}^0_1$
       or $A_1 A_1$ and
       the conventional visible decays (e.g. into bottom quarks) are
       severely suppressed.

\end{abstract}

\maketitle

Because the minimal supersymmetric standard model (MSSM) suffers from
the $\mu$-problem, some non-minimal supersymmetric models have
recently been intensively studied, among which an attractive one is
the nearly minimal supersymmetric standard model (nMSSM)
\cite{Menon,Carena}. This model extends the MSSM by one singlet
superfield $\hat{S}$ with the superpotential~\cite{nMSSM}
\begin{eqnarray}
W & = & W_{MSSM} + \lambda \varepsilon_{ij} \hat{H}_u^i \hat{H}_d^j
\hat{S} + \xi_F M_n^2 \hat{S} \label{Superpotential},
\end{eqnarray}
where $W_{MSSM}$ is the superpotential of the MSSM without the
$\mu$-term, the second term on the right side is the interaction of
the singlet $\hat{S}$ with the Higgs doublets $\hat{H}_u$ and
$\hat{H}_d$, and the last term is the tadpole term.
This superpotential differs from that of the next-to-minimal
supersymmetric standard model (NMSSM)~\cite{NMSSM} in that the
tadpole term of the nMSSM replaces the trilinear singlet term
$\kappa \hat{S}^3$ of the NMSSM.
Due to this
tadpole term, $W$ has no discrete symmetry and so the nMSSM is free
of the domain wall problem suffered by the NMSSM.
The tadpole term also gives rise to a vacuum expectation value (vev)
for the singlet, controlled by $\xi_F M_n^2$.
Though it is SUSY-preserving, this tadpole
term can naturally be of the SUSY breaking scale -- e.g., in the $N=1$
supergravity model with a discrete R-symmetry 
\footnote{Such a discrete R-symmetry is, of course, broken by the SUSY-breaking
soft terms, as well as by the source of SUSY-breaking, which could be a non-zero
constant superpotential induced by the spontaneous breaking in the hidden sector 
or by condensation phenomena. Although this constant in the superpotential 
may be utilized to cancel the cosmological constant, for phenomenological 
study it is usually assumed that all R-symmetry violation is encoded in the 
soft SUSY-breaking terms.} 
such a tadpole term is
generated at a high loop level and thus is naturally
small \cite{nMSSM}.  A nonzero singlet vev at the SUSY breaking scale
generates an effective $\mu$ term from the
$\lambda \varepsilon_{ij} \hat{H}_u^i \hat{H}_d^j \hat{S}$
term with the desired order of magnitude, solving the $\mu$ problem of the
MSSM.
These theoretical virtues motivate further phenomenological study of the nMSSM.
Because of the absence of the trilinear singlet term, the spectrum and
phenomenology of the nMSSM can be quite different from those of the NMSSM.

With the running of the LHC, all low energy supersymmetric models
will soon be put to the test. To explore these models at the LHC, it
is very important to determine the parameter space allowed by
current experiments. In this work we comprehensively examine
experimental constraints on this model from the direct experimental
searches for Higgs bosons and sparticles, the precision electroweak
measurements at LEP/SLD, the cosmic dark matter relic density from
WMAP, and the muon anomalous magnetic moment. We also consider the
theoretical constraints from the stability of the Higgs potential
and the perturbativity of the theory up to the grand unification
scale. After analyzing the allowed parameter space, we discuss
some phenomenology of this model.

\vspace*{0.3cm}


We start our analysis by recapitulating the basics of the nMSSM.
With the superpotential in Eq.~(\ref{Superpotential}), the corresponding
soft-breaking terms are given by \cite{Menon,Barger}
\begin{eqnarray}
V_{\rm soft} & = & V_{MSSM} + \tilde{m}_d^2 |H_d|^2 +
\tilde{m}_u^2 |H_u|^2+ \tilde{m}_S^2 |S|^2 \nonumber \\
&& + (\lambda A_\lambda \varepsilon_{ij} H_u^i H_d^j S + \xi_S M_n^3 S + \mbox{h.c.})
\label{soft}
\end{eqnarray}
where $V_{MSSM}$ contains the soft breaking terms for gauginos and
sfermions in the MSSM, and $\tilde{m}_{u,d,S}$, $A_\lambda$ and
$\xi_S M_n^3$ are soft breaking parameters. Noting that the
tadpole terms do not induce any interactions, one can conclude that,
except for the tree-level Higgs boson masses and the minimization
conditions, the theory is the same as the well-known NMSSM with the
trilinear singlet term set to zero~\cite{Barger}. The nMSSM predicts three
CP-even and two CP-odd neutral Higgs bosons as well as five
neutralinos \cite{Menon,Barger}. The mass of the lightest neutralino
(assumed to be the LSP) arises from the mixing of the singlino with
higgsinos and is given by \cite{lsp-mass}
\begin{eqnarray}
m_{\tilde{\chi}^0_1} \simeq \frac{2\mu \lambda^2 (v_u^2+
v_d^2)}{2\mu^2+\lambda^2  (v_u^2+ v_d^2)}
             \frac{\tan \beta}{\tan^2 \beta+1}
             \label{neutralinomass}
\end{eqnarray}
where $\tan \beta = v_u/v_d$, $\mu = \lambda \langle s \rangle$,
and $v_u$, $v_d$ and $\langle s \rangle$ are the vevs of the
Higgs fields $H_u$, $H_d$ and $S$, respectively.

In our calculations we extend the packages NMSSMTools
\cite{NMSSMTools} and micrOMEGAs \cite{Belanger} to the nMSSM.
We use the modified NMSSMTools to calculate the Higgs boson masses
and their decays including all known radiative corrections.
We use the modified micrOMEGAs to calculate the dark matter relic
density. The parameters relevant to our analysis are $\lambda$,
$A_\lambda$, $\tan \beta$, $m_A \equiv 2( \mu A_\lambda + \lambda
\xi_F M_n^2 )/\sin 2 \beta$, $\mu \equiv \lambda \langle s \rangle$,
$\tilde{m}_S$, the gaugino masses
$M_1$ and $M_2$, and the soft SUSY breaking parameters in the squark/slepton
sectors. We assume all these parameters to be real and specify their values
at the weak scale.

Since we are only interested in the properties of the Higgs bosons
and neutralinos and these properties are affected little by the soft
parameters in the squark/slepton sectors, we specify these parameters
before our scan. Noting that a heavy stop ($\tilde{t}$) is helpful
for the Higgs sector to evade the LEP constraints and a light smuon
($\tilde{\mu}$) is needed in the nMSSM to explain the muon anomalous
magnetic moment $a_\mu$, we assume all the soft breaking parameters
(soft masses and trilinear $A$-parameters) to be 1 TeV for the
$(\tilde{t}, \tilde{b})$ sector and 100 GeV for the
$(\tilde{\nu}_\mu,\tilde{\mu})$ sector. We will briefly
discuss the results with a lower $m_{\tilde{t}}$ and those with
different $m_{\tilde{\mu}}$ when the $a_\mu$ constraint is switched
on.  For the other soft breaking parameters in the squark/slepton
sector, we uniformly set them to be 1~TeV since the considered
constraints are not sensitive to them. Moreover, we assume the grand
unification relation for the gaugino masses, $M_1 = (5g^{\prime
2}/3g^2 ) M_2$. With the above assumptions, we scan over the
remaining seven parameters in the following ranges: $ 0.1 \leq
\lambda \leq 0.7$, $1 \leq  \tan \beta \leq 10$, $ - 1 {\rm~TeV}
\leq A_\lambda  \leq 1 {\rm~TeV} $, $ 50 {\rm~GeV} \leq m_A, \mu,
M_2 \leq 1 {\rm~TeV} $ and $0 < \tilde{m}_S \leq 200 {\rm~GeV}$.
Note that in the nMSSM $\tan \beta>10$ is not allowed by the dark
matter relic density because $m_{\tilde{\chi}^0_1}$ is suppressed by
large $\tan \beta$ (see Eq.~(\ref{neutralinomass})) and a light
$\tilde{\chi}^0_1 $ is difficult to annihilate sufficiently.

In our scan we consider the following constraints:
(1) The dark matter relic density, $0.0945 < \Omega h^2 < 0.1287 $ \cite{dmconstr}.
     We require $\tilde{\chi}^0_1$ to account for this density.
(2) The $a_\mu$ constraint,
    $a_\mu^{exp} - a_\mu^{SM} = ( 29.5 \pm 8.8 ) \times 10^{-10}$ \cite{Miller}.
     We require the  nMSSM contribution to explain the deviation at the $2 \sigma$ level.
(3) The LEP bound on invisible $Z$ decay, $\Gamma(Z \to
\tilde{\chi}^0_1 \tilde{\chi}^0_1) < 1.76$ MeV \cite{Amsler}
(we also apply this bound to the decay $Z \to h_1 A_1$ with $h_1$ being
the lightest CP-even Higgs boson);
the LEP-II upper bound on $\sigma(e^+e^- \to
\tilde{\chi}^0_i \tilde{\chi}^0_j)$, which is $10^{-2}~{\rm pb}$
for $i=1$, $j>1$ and $10^{-1}~{\rm pb}$ for $i,j>1$
(summed over $i$ and $j$) \cite{Abdallah};
and the lower mass bounds on sparticles from direct searches at LEP and
the Tevatron \cite{Amsler}. (4) Constraints from the direct search for
Higgs bosons at LEP-II, which limit all possible channels for the production
of Higgs bosons.
(5) Constraints from precision electroweak observables such as
           $\rho_{lept}$, $\sin^2 \theta_{eff}^{lept}$, $m_W$ and $R_b$.
(6) The perturbativity of the nMSSM up to the grand
           unification scale and the stability of the Higgs potential
           which requires that the physical vacuum of the Higgs potential is
           the true minimum of the potential.

The above constraints have been encoded in NMSSMTools
\cite{NMSSMTools}, except for (5). In Ref.~\cite{Cao} we extended the code by
adding (5); here we extend all these constraints to the nMSSM
scenario. Since the hadronic contribution to $a_\mu^{SM}$ remains under
discussion~\cite{Miller}, in the following we will present results both
with and without considering the $a_\mu$ constraint.

\vspace*{0.3cm}


Our scan sample is 2.5 billion random points in the parameter space
given above. With all the constraints except $a_\mu$, only about 6
thousand points survive. 
This is mainly because the dark matter relic
density stringently constrains the mass
and couplings of the LSP, and consequently, only a small 
portion of the parameter space is allowed (as shown explicitly in 
Figs.1-4 below). 
Among the surviving points, about $60\%$ are characterized by 
$m_{A_1} \ge m_Z $ and $30$ GeV $\leq m_{\tilde{\chi}^0_1} \leq 37$ GeV, 
in which $\tilde{\chi}^0_1$ mainly annihilates through $Z$-boson
exchange to give the required dark matter relic density \cite{Menon}.  
For most of the other points, both
$\tilde{\chi}^0_1$ and $A_1$ are predominantly singlet-like with
roughly $2 m_{\tilde{\chi}^0_1} \sim m_{A_1}$, in which 
the exchange of a light $A_1$ is the main annihilation channel of
$\tilde{\chi}^0_1$ (this annihilation channel, similar to the case 
of Table III in \cite{Gunion} for the NMSSM, has not been studied 
for the nMSSM in the literature).

If we switch on the $a_\mu$ constraint, the surviving points are
reduced to 1.4 thousand, among which about 60\% (70\%) satisfy
$m_{A_1} < 60$ GeV  ($m_{A_1} < m_Z$). This indicates that the
current $a_\mu$ measurement can constrain the model stringently,
and that it favors
LSP annihilation through exchange of a light $A_1$ rather than
a $Z$-boson. Note that in the above we fixed
$m_{\tilde{\mu}} = 100$~GeV. If we increase $m_{\tilde{\mu}}$, even
fewer scan points will survive. For
example, raising $m_{\tilde{\mu}}$ to 150~GeV eliminates about half
the remaining points with $m_{A_1} < m_Z$ and nearly all
the points with $m_{A_1} \ge m_Z$.
For $m_{\tilde{\mu}}= 200$~GeV, none of our scan points survive the
$a_\mu$ constraint, which implies that smuons must be lighter than
about 200~GeV. This conclusion is unique to the nMSSM. The underlying
reason is that in SUSY models, the leading chargino/neutralino
contribution to $\delta a_\mu$ is proportional to $\tan
\beta/m_{\tilde{\mu}}^N$ with $N \ge 2$ \cite{Martin,Gunion}.
In the MSSM or NMSSM, $\tan \beta$ can be quite large \cite{Cao} and thus
$m_{\tilde{\mu}}$ is not stringently constrained by $a_\mu$; but in the
nMSSM, $\tan \beta$ is bounded from above ($\lsim 10$) by the dark matter
relic density and hence $m_{\tilde{\mu}}$ must be light.

\begin{figure}[htb]
\epsfig{file=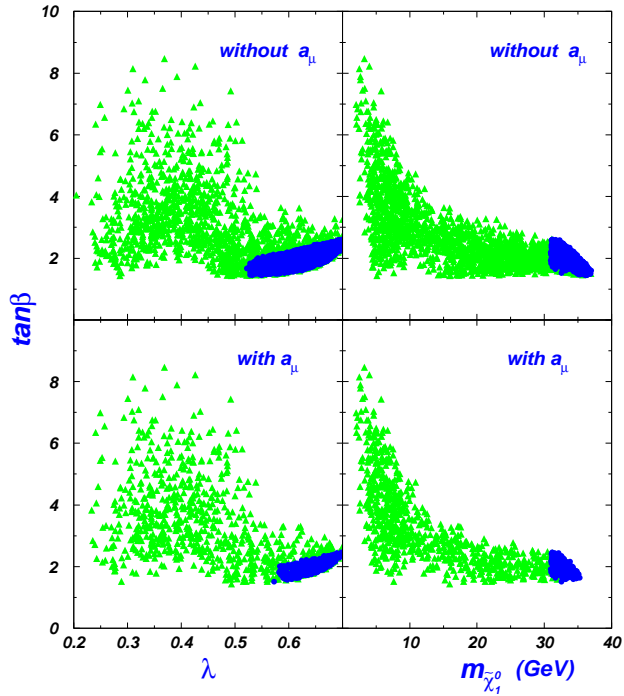,width=8.5cm}
\vspace*{-0.3cm}
\caption{Parameter scan points allowed by current experiments.
         The dark bullets (light triangles) correspond to $m_{A_1} \geq m_Z$
         ($m_{A_1} < m_Z$), with $\tilde{\chi}^0_1$ mainly annihilating through
         exchanging a $Z$-boson (a light $A_1$) to give the
         required dark matter relic density. }
\end{figure}
\begin{figure}[htb]
\epsfig{file=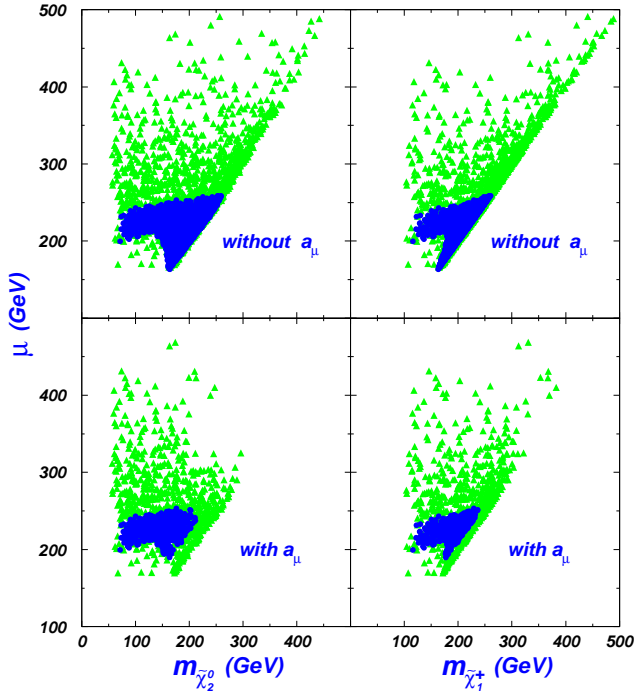,width=8.5cm}
\vspace*{-0.3cm}
\caption{Same as Fig.~1, but showing dependence on $\mu$,
$m_{\tilde \chi_2^0}$, and $m_{\tilde \chi_1^+}$.}
\end{figure}

In Figs.~1 and 2 we display the surviving points as functions of
various parameters. As we pointed out before, the strongest
constraint come from the dark matter relic density. These figures
show that if only the $Z$ exchange is responsible for the density
(dark bullets), the parameters $\lambda$, $\tan \beta$, $\mu$,
$m_{\tilde{\chi}^0_{1,2}}$ and $m_{\tilde{\chi}^+_1}$ are all
constrained in quite narrow ranges; however, if the light $A_1$
exchange is considered (light triangles), the allowed parameter
ranges are significantly more spread out.

Note that we checked by using the modified NMSSMTools that $b$--$s$
transitions such as $b \to s \gamma$ and $B_s^0$--$\bar{B}_s^0$ mixing
do not impose any meaningful constraints on the surviving samples in
the case of no squark flavor mixings. We also checked that for
samples with $m_{A_1} \lesssim 10$ GeV, the branching ratio for
$\Upsilon \to \gamma A_1$ is less than its experimental upper bound of
$1 \times 10^{-4}$~\cite{Rosner}.

The following additional comments are in order.  (i) We call $\tilde\chi^0_1$
singlino-like when the coefficient of the singlino component in
$\tilde\chi^0_1$ is larger than $1/\sqrt{2}$ (so its square is
larger than $1/2$).  In general the
higgsino or gaugino components in $\tilde\chi^0_1$ are not negligible
even when $\tilde\chi^0_1$ is singlino-like.
In fact, it is the higgsino components in $\tilde\chi^0_1$ that are
mainly responsible for the LSP annihilation coupling discussed above. For
annihilation via $Z$ exchange, the typical coefficient of the $H_u$-type
higgsino component in $\tilde\chi^0_1$ is 0.4, and we checked that
as the higgsino components increase, the coupling of
$\tilde{\chi}_1^0 \tilde{\chi}_1^0 Z$ increases and consequently the
relic density drops \cite{Menon}. For annihilation via $A_1$ exchange,
the typical coefficient for the higgsino component in
$\tilde{\chi}_1^0$ is 0.2 and for the doublet-Higgs component in
$A_1$ is 0.15. Since the couplings of $\tilde\chi^0_1 \tilde\chi^0_1 A_1 $
and $A_1 \bar{f} f$ (where $f$ is a light fermion) are suppressed,
annihilation via $A_1$ exchange is too weak to produce the
required relic density except in the funnel region $2
m_{\tilde{\chi}_1^0} \sim m_{A_1}$~\cite{Gunion}.  (ii)  The
requirement $\tilde{m}_S < 200$ GeV in our scan is taken from
Ref.~\cite{Menon} which studied electroweak baryogenesis. We
checked that a larger range of $\tilde{m}_S$ does not change our
conclusions qualitatively; it only increases the number of surviving samples
with $m_{A_1} > m_Z$ and raises the upper bound of $m_{A_1}$.
We did not impose the requirement of successful electroweak baryogenesis
in our scan since there exist other ways to explain the
origin of the matter-antimatter asymmetry in the
Universe~\cite{Menon}.  (iii) A heavy $\tilde{t}$ is not necessary in
the nMSSM for the SM-like Higgs boson to evade
the LEP bound since the Higgs boson mass can be enhanced at tree level by
a contribution from $\lambda$~\cite{NMSSM}. We checked that a lighter
$\tilde{t}$ in our scan does not change our conclusions; it only
decreases the number of the surviving points.
 \vspace*{0.3cm}


As shown above, if we consider all the constraints including $a_\mu$
with $m_{\tilde{\mu}}=100$ GeV, the mass spectrum is limited to the
following ranges:
$1 {\rm GeV} \lsim  m_{A_1} \lsim  250 {\rm GeV}$, $30 {\rm
GeV} \lsim m_{h_1} \lsim 140 {\rm GeV}$, $70 {\rm GeV} \lsim m_h
\lsim 145 {\rm GeV}$ ($h$ is the lightest doublet-dominant CP-even
Higgs boson, usually called the SM-like Higgs boson; in some
cases $h$ can be $h_1$), $1 {\rm GeV} \lsim
m_{\tilde{\chi}^0_1} \lsim 37 {\rm GeV}$, $50 {\rm GeV}\lsim
m_{\tilde{\chi}^0_2} \lsim 300 {\rm GeV}$ and $105 {\rm GeV}\lsim
m_{\tilde{\chi}^+_1} \lsim 400 {\rm GeV}$. In such allowed mass
ranges, the phenomenology of the Higgs bosons and sparticles may be
quite peculiar and different from the MSSM. A comprehensive
study of the phenomenology of this model at the Tevatron and the
LHC is beyond the scope of this paper; instead we present
the following brief discussion.

First, consider the Higgs bosons. The dominant decay of $A_1$
may be either $\tilde{\chi}^0_1 \tilde{\chi}^0_1$ or $b \bar{b}$ for
$m_{A_1} > 2 m_b$, and we checked that the $A_1 \tilde{\chi}_1^0
\tilde{\chi}_1^0 $ interaction is mainly induced by the higgsino
components of $\tilde{\chi}_1^0$ and/or the doublet components of
$A_1$ \cite{NMSSMTools}.
The dominant decay mode of the SM-like Higgs boson $h$ can be any
of the following: $h \to
\tilde{\chi}^0_1 \tilde{\chi}^0_1, \tilde{\chi}^0_1
\tilde{\chi}^0_2, A_1 A_1, h_1 h_1 $. In Fig.~3 we show the branching
ratios of $h$ and $A_1$ decays into the LSP and $b \bar{b}$.
For $m_{A_1}\gsim 60$ GeV, both $h$ and $A_1$ can
decay predominantly into LSP pairs and the decay into $b \bar{b}$
is strongly suppressed.
Since the branching ratio of $h \to b \bar{b}$ is suppressed below
$10\%$ in most of the allowed parameter space due to the presence of
new competing decay modes, conventional
searches for a light SM-like CP-even Higgs boson at the LHC will be
quite hopeless. For example, our results indicate that for about
$60\%$ of the surviving points, the final decay products of $h$ are
two or four LSPs. For these points, weak boson fusion with $h \to$ invisible
is a good search channel~\cite{Cavalli}. Our results also indicate
that for about $16\%$ of the surviving points, $h$ decays predominantly
to $4 b$. In this case, $W/Z h$ production may be a good channel to
detect $h$~\cite{Cheung}.

\begin{figure}[htb]
\epsfig{file=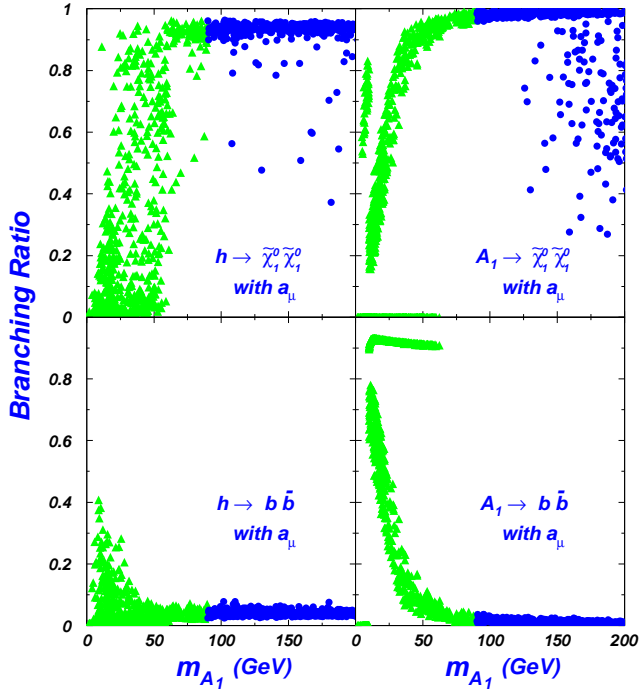,width=8.5cm} \vspace*{-0.3cm} \caption{Same as
Fig.~1, but for the branching ratios of $h$ and $A_1$
         versus $m_{A_1}$ with all constraints imposed including $a_\mu$.
         The effect of the $a_\mu$ constraint is to reduce the number
         of surviving points, as shown in Figs.~1 and 2.}
\end{figure}

Second, consider smuon production at the LHC. If we take the $a_\mu$ constraint
seriously, the smuon should be lighter than 200 GeV, which implies that it
would be copiously produced either directly or from cascade decays of
other sparticles at the LHC, and should be visible at the LHC. We
checked that due to the non-negligible gaugino component of
$\tilde\chi^0_1$, the decay width of $\tilde{\mu} \to \mu
\tilde{\chi}_1^0$ is about several MeV, so the decay length of smuon
is not macroscopic (for a heavy charged particle with macroscopic
decay length, its signals at the LHC may be quite special \cite{tata}).

\begin{figure}[htb]
\epsfig{file=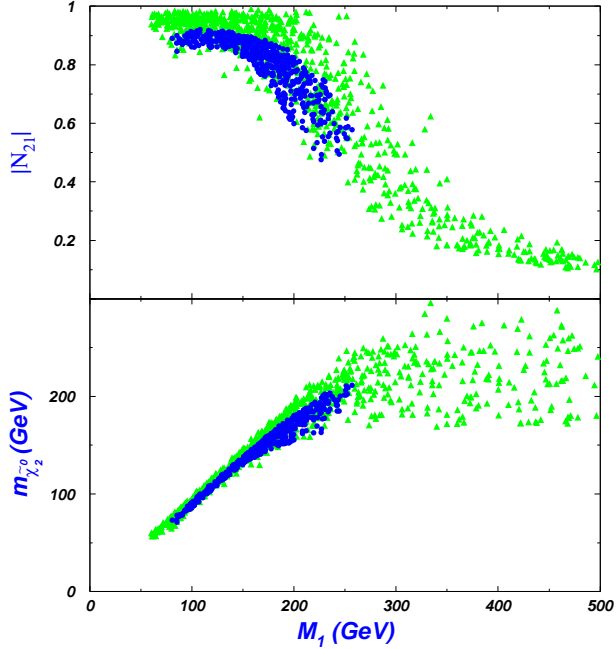,width=8.5cm} \vspace*{-0.3cm} 
\caption{Same as Fig.~1,  but showing dependence on $M_1$, 
$m_{\tilde{\chi}^0_2}$ and $|N_{21}|$ with all constraints 
imposed including $a_\mu$. $N_{21}$ denotes
the coefficient of the bino component in $\tilde{\chi}^0_2$.}
\end{figure}

Third, consider the next-to-lightest neutralino $\tilde{\chi}^0_2$
at the LHC. As shown in Fig.4, $\tilde{\chi}^0_2$ is bino-like for
$m_{\tilde{\chi}^0_2} \lsim 150$ GeV, and may be higgsino-like for
$m_{\tilde{\chi}^0_2} > 150$ GeV. We checked that the main decay
products of $\tilde{\chi}_2^0$ can be $\tilde{\chi}^0_1 h_1$,
$\tilde{\chi}^0_1 A_1$, $\tilde{\chi}^0_1 h$, or $\tilde{\chi}^0_1
Z$ and there is a large portion of the surviving samples in which
$\tilde{\chi}^0_2  \to \tilde{\chi}^0_1 A_1 (h_1,h) \to
\tilde{\chi}^0_1 \tilde{\chi}^0_1 \tilde{\chi}^0_1 $. If
$\tilde{\chi}^0_2$ is the NLSP and it mainly decays into $3
\tilde{\chi}^0_1$,  it will be copiously produced from the cascade
decay of squarks and gluinos \cite{Carena}, and can be easily
mistaken as the LSP. From Fig.4 one can also learn that the gaugino
mass $M_1$ is bounded from 50 GeV to 500 GeV. This implies by the
gaugino mass unification relation that the gluino mass varies from 
about 300 GeV to 3 TeV, which could be accessible at the LHC. 
\vspace*{0.2cm}

In summary, we examined the current experimental constraints on the nMSSM.
We found that the parameter space of this model is stringently constrained
by current experiments, and in the allowed parameter space
the phenomenology of this model may be quite peculiar.
Such tightly constrained parameter space could make this model
readily tested (either verified or excluded) at the LHC.
In addition, since in this model the dark matter particle
is constrained in a narrow mass range, the astrophysical dark matter
experiments may also be able to cast some light on this model.
\vspace*{0.2cm}

We thank Prof. C. Hugonie for useful discussions about NMSSMTools and microMEGAS.
This work was supported in part by the Natural Sciences and
Engineering Research Council of Canada, by the China
NSFC under grant No.
10505007, 10821504, 10725526 and 10635030, and by HASTIT under grant
No. 2009HASTIT004.



\begin{thebibliography}{99}
\bibitem{Menon}
  A.~Menon, {\it{et. al.}},
  Phys.\ Rev.\  D {\bf 70}, 035005 (2004);
     V.~Barger, {\it{et. al.}},
  Phys.\ Lett.\  B {\bf 630}, 85 (2005).

\bibitem{Carena}
C.~Balazs, {\it{et. al.}},
  JHEP {\bf 0706}, 066 (2007).

\bibitem{nMSSM}
  C.~Panagiotakopoulos, K.~Tamvakis,
  Phys.\ Lett.\  B {\bf 446}, 224 (1999);
   Phys.\ Lett.\  B {\bf 469}, 145 (1999);
    C.~Panagiotakopoulos, A. Pilaftsis
    Phys.\ Rev.\  D {\bf 63}, 055003 (2001);
    A.~Dedes, {\it{et. al.}}, Phys.\ Rev.\  D {\bf 63}, 055009 (2001);
    P. Fayet, \NPB90, 104 (1975).

\bibitem{NMSSM}
J.~R.~Ellis, {\it{et. al.}},
Phys.\ Rev.\ D {\bf 39} (1989) 844;
%
M.~Drees,
Int.\ J.\ Mod.\ Phys.\ A {\bf 4} (1989) 3635.

\bibitem{Barger}
  V.~Barger, {\it{et. al.}},
  Phys.\ Rev.\  D {\bf 73}, 115010 (2006).


\bibitem{lsp-mass}  S. Hesselbach, {\it{et. al.}}, arXiv:0810.0511v2 [hep-ph].

\bibitem{NMSSMTools}
  U.~Ellwanger, {\it{et. al.}},
  JHEP {\bf 0502}, 066 (2005).

\bibitem{Belanger}
G. Belanger, F. Boudjema, C. Hugonie, A. Pukhov, A. Semenov, JCAP 0509, 001 (2005);
G. Belanger, F. Boudjema, A. Pukhov, A. Semenov, Comput. Phys. Commun. 176, 367 (2007);
C. Hugonie, G. Belanger, A. Pukhov, JCAP0711, 009 (2007).

\bibitem{dmconstr}
C.~L.~Bennett {\it et al.}, Astrophys.\ J.\ Suppl.\ {\bf 148} (2003)
1; D.~N.~Spergel {\it et al.}, Astrophys.\ J.\ Suppl.\ {\bf 148}
(2003) 175.

\bibitem{Miller}
  J.~P.~Miller, {\it{et. al.}},
  Rept.\ Prog.\ Phys.\  {\bf 70}, 795 (2007).

\bibitem{Amsler}
  C.~Amsler {\it et al.},
  Phys.\ Lett.\  B {\bf 667}, 1 (2008).

\bibitem{Abdallah}
  J.~Abdallah {\it et al.},
  Eur.\ Phys.\ J.\  C {\bf 31}, 421 (2004);
 G.~Abbiendi {\it et al.},
  Eur.\ Phys.\ J.\  C {\bf 35}, 1 (2004).

\bibitem{Cao}
  J. J.  Cao, J. M. Yang, \JHEP0812, 006 (2008); \PRD78, 115001 (2008).

\bibitem{Gunion}
  J.~F.~Gunion, {\it{et. al.}},
  Phys.\ Rev.\  D {\bf 73}, 015011 (2006).



\bibitem{Martin}
  S.~P.~Martin and J.~D.~Wells,
  Phys.\ Rev.\  D {\bf 64}, 035003 (2001).


\bibitem{tata} M. Drees, X. Tata, \PLB252, 695 (1990).

\bibitem{Cavalli}
  D.~Cavalli {\it et al.},
  arXiv:hep-ph/0203056;
  H.~Davoudiasl {\it et al.},
  Phys.\ Rev.\  D {\bf 71}, 115007 (2005).

\bibitem{Cheung}
  K.~Cheung {\it et al.},
  Phys.\ Rev.\ Lett.\  {\bf 99}, 031801 (2007);
  M.~Carena {\it et al.},
  JHEP {\bf 0804}, 092 (2008).

\bibitem{Rosner}
  J.~L.~Rosner {\it et al.}  [CLEO Collaboration],
  Phys.\ Rev.\  D {\bf 76}, 117102 (2007).

\end{thebibliography}
\end{document}